\documentclass[usenatbib]{basi}

\usepackage{epstopdf}
\usepackage{amsmath}  
\usepackage{amsfonts} 
\usepackage{graphicx} 
\usepackage{amsthm}
\usepackage{float}
\usepackage{mathtools}
\usepackage{empheq}
\usepackage{natbib}

\usepackage[T1]{fontenc}
\usepackage[british]{babel}
\usepackage[varg]{txfonts}
\usepackage{rotating}
\usepackage{dcolumn}

\begin{document}
\title{Theoretical Model for Time Evolution of an Electron Population under Synchrotron Loss}

 
        
\author[S. Malik 2015]%
       {Siddharth Malik$^1$\thanks{email: \texttt{siddharthmalik104@gmail.com}}\\
       $^1$Department of Earth and Space Sciences, Indian Institute of Space Science and Technology, Thiruvananthapuram-695547, India\\
       $^2$Semiconductor Laboratory, Department of Space, Government of India, Mohali-160071, India}

\pubyear{2015}
\volume{00}
\pagerange{\pageref{firstpage}--\pageref{lastpage}}        
        
\maketitle

\label{firstpage}

\begin{abstract}

Many astrophysical sources radiate via synchrotron emission from relativistic electrons. The electrons give off their kinetic energy as radiation and this radiative loss modifies the electron energy distribution. An analytical treatment of this problem is possible in asymptotic limits by employing the continuity equation. In this article, we are using a probabilistic approach to obtain the analytical results. The basic logic behind this approach is that any particle distribution can be viewed as a probability distribution after normalizing it (as is done frequently in statistical mechanics with ensembles containing very large number of particles). We are able to reproduce the established results from our novel approach. Same approach can be applied to other physics problems involving spatial or temporal evolution of distribution functions. 

\end{abstract}
\section{Introduction}

Synchrotron radiation is emitted by relativistic charged particles circling magnetic field
lines. Highly relativistic particles are abundant in our universe and the radiation detected
from a range of astrophysical sources originate due to this process. Supernova remnants,
pulsars, Gamma Ray Bursts and large-scale jets emanating from certain galaxies are some
of the examples. In this article, we are describing the time evolution of a population of
electrons emitting synchrotron radiation. We used probability theory to arrive at the results
conventionally derived using the continuity equation. These results are widely used in modeling astrophysical sources. Our approach is generic and can be used for similar physical
problems.

Total energy of a relativistic electron can be expressed as $E = \gamma m_e c^2$ where $m_{e}c^2$ is the rest energy and $\gamma$ is the Lorentz factor ( $\sqrt[2]{1-v^2/c^2}$ ) of an electron moving with velocity $v$. The synchrotron power $P_{syn}$ depends on the energy E (equivalently lorentz factor $\gamma$) and the larmor radius of the electron. As the electron radiates, its energy (equivalently $\gamma$) reduces. Time evolution of the Lorentz factor of a synchrotron radiating electron of mass $m_e$, charge $e$ and initial lorentz factor $\gamma_0$ in the presence of a magnetic field density B can be derived as follows (see appendix for the derivation)

\begin{equation}
\gamma(t) = \frac{\gamma_0}{1 + A\gamma_0t} \hspace{10 mm}\hspace{10 mm} A = \frac{\sigma_{t}B^2}{6\pi m_{e}c} 
\end{equation}
where $\sigma_{t}$ is the Thompson's Cross section.

Instead of monoenergetic electrons, realistic systems have electrons of a certain energy distribution. Most astrophysical systems emitting synchrotron radiation do not have sufficient particle density to achieve thermal equilibrium. Hence instead of a relativistic Maxwellian, the energy distribution is expected to be a `power-law' of the form :
\begin{equation}
n(\gamma)d\gamma = K_e\gamma^{-p}d\gamma   \textrm{    where   } \gamma_m < \gamma < \gamma_u
\end{equation}
where $\gamma_m$ and $\gamma_m$ are the minimum and maximum limits of the distribution respectively, $K_e$ is an arbitrary constant depending on the physical parameters of a given system. This is also called the `non-thermal' distribution. This distribution does not remain the same over time due to synchrotron loss suffered by each electron. However, the time evolution of the electron population not only depends on the energy loss but also on the injection of fresh particles into the system. Our aim is to obtain the distribution $n'(\gamma(t))$ at any given time t. 

We assume a constant magnetic field B and two types of injection : (i) One Shot injection and (ii) Continuous Injection. We assume that the fresh particles that are injected in the system always follow the non-thermal powerlaw given in equation 2.

\section{One Shot Injection}

One shot injection means that at t=0 we inject electrons in the system which follow the power-law distribution in equation 2. If we divide equation 2 by the total number of particles in the system N, it becomes a probability distribution :
\begin{equation}
p(\gamma_0)d\gamma_0 = K\gamma_0^{-p}d\gamma_0   \textrm{    where   } \gamma_m < \gamma_0 < \infty
\end{equation}
We have assumed that $\gamma_u \rightarrow \infty$. $K=K_e/N$ and $\gamma_m$ is the minimum lorentz factor of the distribution. $p(\gamma_0)=$ is the probability that an electron has an initial Lorentz factor between $\gamma_0$ and $\gamma_0$ + $d\gamma_0$ 

If we substitute $\gamma_0$ as $\infty$ in equation 1, then $\gamma(t)$ will tend to $\frac{1}{At}$. Therefore, even if the maximum lorentz factor in infinite initially, it will to tend to some finite value say $\gamma_u$ at a later time t.

The system at a later time t can be represented as,
\begin{equation}
p(\gamma)d\gamma = K\gamma^{-p}d\gamma   \textrm{    where   } \gamma_m < \gamma < \gamma_u
\end{equation}
$K=K_e/N$; $p(\gamma)=$ probability that a electron has lorentz factor between $\gamma$ and $\gamma + d\gamma$ at a given time $t$.

Let us consider two random variables:
\begin{enumerate}
\item $\gamma_0$ which represents the Lorentz factor of an electron at time $t=0$
\item $\gamma$ which represents the Lorentz factor of an electron at some time $t$
\end{enumerate} 
In order to find $p(\gamma)$, we can use the theorem for transformation of random variables from probability theory \citep*{2}.

\newtheorem{theorem 1}{Theorem}
\begin{theorem 1}
Suppose that $X$ a continuous random variable with probability distribution $f(x)$. Let $y = u(x)$ define a one-one relationship correspondence between the values of $X$ and $Y$ so that the equation $y = u(x)$ can be uniquely solved for $x$ to obtain $x = w(y)$. Then, the probability distribution of $Y$ is given is:
\begin{equation}
g(y) = f(w(y))|J|
\end{equation} 
where $J = dw/dy$ and is called the \emph{jacobian} of the transformation
\end{theorem 1}

In our case $x=\gamma_0$, $f(x)=p^{'}(\gamma_0)$, $y=\gamma$ and $g(y)=p(\gamma)$, $u(x)$ is given by equation 2. So, using Theorem 1, we can write,

\begin{equation}
p(\gamma) = p^{'}(\gamma_0(\gamma))\frac{d\gamma_0}{d\gamma} = K\frac{\gamma^{-p}}{{(1-A\gamma t)}^{-p+2}}
\end{equation}
or,
\begin{equation}
p(\gamma)d\gamma = K\frac{\gamma^{-p}}{{(1-A\gamma t)}^{-p+2}}d\gamma \hspace{10 mm}where\hspace{10 mm}\gamma^{'}_m < \gamma < \gamma^{'}_u
\end{equation}

At a later time $t$, minimum and maximum values of $\gamma$ have changed which can be found out by substituting initial minimum ($\gamma$) and maximum ($\infty$) values in equation 2.
\begin{equation}
\gamma^{'}_m = \frac{\gamma_m}{1+A\gamma t} \hspace{15 mm} \gamma_c \equiv \gamma^{'}_u = \frac{1}{At}
\end{equation}

Note, that even if at $t=0$, maximum possible Lorentz factor for an electron was infinite, after a time t, maximum possible Lorentz factor $= \gamma_c$. No electron can have a Lorentz factor greater than $\gamma_c$ at a given time t.

Equation 7 gives us the probability distribution. For obtaining the particle distribution function, we have to multiply the probability distribution function by $N$ (total number of particles).
\begin{equation}
n(\gamma)d\gamma = K_e\frac{\gamma^{-p}}{{(1-A\gamma t)}^{-p+2}}d\gamma \hspace{10 mm}where\hspace{10 mm}\gamma^{'}_m < \gamma < \gamma_c
\end{equation} 

where $K_e=NK$. So, the particle distribution function at time $t$ for one shot injection is given by equation 9. Figure 1 shows both initial and time evolved distribution functions. The graph is plotted between $\gamma_m$ and $\gamma_c$.  

\begin{figure}

\includegraphics{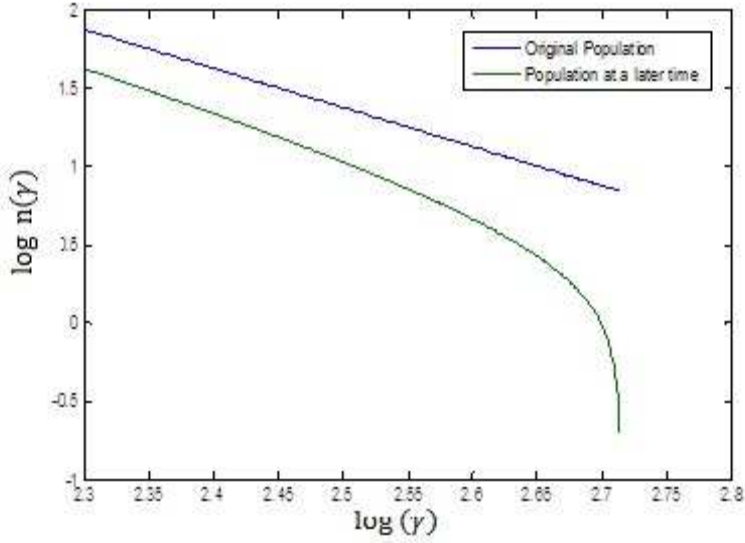}
\caption{Initial and Time Evolved Particle Distribution for one shot injection given by equation 2 and equation 9 respectively}

\end{figure}

\section{Continuous Injection}

In the case of Continuous Injection, fresh electrons continuously get added and make up for the loss in energy space. Now, total number of particles will also vary. We can treat Continuous Injection as a series of one shot injections. For simplicity, let's assume that at each instant of time we inject $N_0$ number of particles and each injection follows equation 1 with same $p, \gamma_m$ and $\gamma_u = \infty$.
\begin{equation}
n(\gamma_0)d\gamma_0 = K_e\gamma_0^{-p}d\gamma_0   \textrm{    where   } \gamma_m < \gamma_0 < \infty
\end{equation} 
$K_e$ is determined from the equation 
\begin{equation}
\int_{\gamma_m}^\infty n(\gamma_0)d\gamma_0 = N_0
\end{equation}
\begin{equation}
N=N_0t
\end{equation}

Now, again consider two random variables $\gamma_0$ and $\gamma$ in exactly the same manner as described for the one shot injection case. Each electron in the system was injected during a particular injection and each injection evolves independent of other injections. If we look at any one injection, it evolves in a manner similar to what we discussed in previous section for one shot injection. The overall evolution can be viewed as a summation or integration over all injections. If we are looking at time t we can assign two other variables to each electron $t_1$ and $t_2$, $t_1$ is the time when the electron was injected and $t_2$ is the time elapsed since its injection or the time for which the electron has been there in the system. Both $t_1$ and $t_2$ varies from 0 to $t$ such that $t_1 + t_2=t$.

Define:

\begin{description}
\item [$p(\gamma,t_2)$] Probability that a random electron has a Lorentz factor $\gamma$ at time $t$ and has been there in the system for time $t_2$.
\item [$p(\gamma)$] Probability that a random electron has Lorentz factor $\gamma$ at time $t$.
\item [$p(t_2)$] Probability that a random electron has been there in the system for time $t_2$.
\item [$p(\gamma/t_2)$] Probability that a random electron has a Lorentz factor $\gamma$ given that it has been there in the system for time $t_2$.
\end{description}

From definition of conditional probability distribution, we know that:
\begin{equation}
p(\gamma,t_2) = p(t_2)p(\gamma/t_2)
\end{equation} 
As the injection is identical at every instant, $p(t_2)$ is a uniform distribution:
\begin{equation}
p(t_2)=
\begin{dcases}
\ \frac{1}{t} & \text{if } 0 \leq t_2 \leq t \\
0             & \text{otherwise }
\end{dcases}
\end{equation}

$p(\gamma/t_2)$ is the probability that a random electron has a Lorentz factor $\gamma$ given that it has been there in the system for time $t_2$. If we know the initial distribution and time elapsed, then, the final distribution is given in a same way as for one shot injection case (Once we fix $t_2$, we are talking about a particular injection, so the problem is identical to one shot injection case). Therefore,
\begin{equation}
p(\gamma/t_2) = K\frac{\gamma^{-p}}{{(1-A\gamma t)}^{-p+2}} \hspace{10 mm}where\hspace{10 mm} 0 \leq t_2 \leq t
\end{equation}

where, $K=K_e/N_0$, putting equation 14 and 15 in equation 13:
\begin{equation}
p(\gamma,t_2) = \frac{K}{t}\frac{\gamma^{-p}}{{(1-A\gamma t)}^{-p+2}} \hspace{10 mm} 0 \leq t_2 \leq t
\end{equation}

So, far we have not imposed any condition on $\gamma$. Now from definition of marginal distribution function we know that:
\begin{equation}
p(\gamma) = \int p(\gamma,t_2)dt_2
\end{equation}

Substituting equation 16 in equation 17:
\begin{equation}
p(\gamma) = \int_0^t \frac{K}{t}\frac{\gamma^{-p}}{{(1-A\gamma t)}^{-p+2}} dt_2 \\
\end{equation}
\begin{equation}
p(\gamma) = \frac{K}{A}\frac{\gamma^{-p+1}}{(p-1)t}\left[1 - {(1 - \frac{\gamma}{\gamma_c})}^{p-1}\right]
\end{equation}
$\gamma_c$ and $\gamma^{'}_m$ are defined by equation 8.

Depending upon $\gamma_c$ and $\gamma_m$, we can have two cases: slow cooling ($\gamma_m < \gamma_c$) and fast cooling ($\gamma_m > \gamma_c$). 

The loss becomes considerable only for those electrons for which the radiative time scale is less than the age of the system ($t_{rad} < t$) because only those electrons will dissipate energy at a faster rate than the age of the system. Radiative time scale is given as:
\begin{equation}
t_{rad} \equiv \frac{E}{dE/dt}
\end{equation}

So, loss becomes considerable for those electrons for which
\begin{equation}
t_{rad} \equiv \frac{E}{dE/dt} \leq t
\end{equation}

From equation 1, we can easily derive:
\begin{equation}
\frac{d\gamma}{dt} = \frac{A{\gamma_0}^{2}}{1+A\gamma_{0}t} = A{\gamma}^2
\end{equation}

Using above equation and the relation $E=\gamma m_e c^2$, we can easily derive the following relation:
\begin{equation}
\frac{dE}{dt} = \frac{d(\gamma m_e c^2)}{dt} 
              = \frac{d\gamma}{dt}m_e c^2 
              = A{\gamma^2}m_e c^2  
\end{equation}
Substituting above equation in equation 21 we get,
\begin{equation}
\frac{\gamma m_e c^2}{A{\gamma}^{2}m_e c^2} < t \hspace{5 mm} or, \hspace{5 mm} 
\gamma > \frac{1}{At} = \gamma_c 
\end{equation}

Hence, Equation 21 is equivalent to equation 24. Both are just different representations of the same concept. Both implies that the energy loss is greater for electrons having a lorentz factor greater than $\gamma_c$, we expect the power law to be steeper in the range $\gamma > \gamma_c$. 

Therefore, if $\gamma_m < \gamma_c$, only a small fraction of electrons in the system will be cooling dominated. Hence we refer it to as \emph{slow cooling}. However, if $\gamma_m > \gamma_c$, a large fraction of electrons is cooling dominated and hence we refer it to as \emph{fast cooling}. The electron distribution at a given time is different for both cases. Hence, we have discussed them separately in coming subsections.

\subsection{Slow Cooling}
The expression in equation 19 is valid only for $\gamma_m < \gamma < \gamma_c$. From equation 8,we know that the maximum attainable lorentz factor for a electron at a time t is given by $\gamma_c = 1/At$. A continuous injection can be seen as a superimposition of various sequentially injected one shot injections. Hence, we can apply the results derived for one shot injection on a particular sub-injection of the continuous injection. 

Consider a continuous injection started at $t=0$ and current time as $t$. Now, take a one shot injection injected at time $t_1$ into the system. Then, any electron which was a part of this injection has been inside the system for $t_2 = t- t_1$ seconds. Therefore, the maximum attainable lorentz factor for such a electron is $1/At_2$.

Suppose we want to calculate $p(\gamma)$ for some $\gamma \geq \gamma_c=1/At$. Only those electrons can attain this value whose maximum attainable lorentz factor is greater than the value of $\gamma$ at which we are calculating $p(\gamma)$ i.e. electrons for which
\begin{equation}
\frac{1}{At_2} > \gamma
\end{equation}
or,
\begin{equation}
t_2 < \frac{1}{A\gamma}
\end{equation}

Therefore, for calculating $p(\gamma)$ for $\gamma > \gamma_c$, upper limit of the integral in the equation 18 will change to $1/A\gamma$ instead of $t$.

Hence, for $\gamma > \gamma_c$
\begin{equation}
p(\gamma) = \int_0^{\frac{1}{A\gamma}} \frac{K}{t}\frac{\gamma^{-p}}{{(1-A\gamma t)}^{-p+2}} dt_2 \\
\end{equation}
\begin{equation}
p(\gamma) = \frac{K}{A(p-1)t}\gamma^{-(p+1)} \hspace{10 mm} \gamma > \gamma_c
\end{equation}

Equation 19 and 28 combined gives $p(\gamma)$ for $\gamma_m \leq \gamma < \infty$, but there will be some electrons which will have Lorentz factor below $\gamma_m$ also (electrons having initial Lorentz factor just above $\gamma_m$ will lose energy to go below $\gamma_m$. For these electrons also, we will have a different expression. The minimum Lorentz factor attainable at time $t$ is $\gamma_m^{'}$, which is given by given by equation 8. But this is the minimum for the very first injection injected at $t = 0$. For electrons belonging to any later injection (which were injected at time $t_1 > 0$ and has been there for time $t_2 < t$), the minimum attainable Lorentz factor is $\gamma_m/(1 + A\gamma_mt_2)$. So, if we are looking at a $\gamma < \gamma_m$, only those electrons have a non-zero probability of having this Lorentz factor which were injected such that the minimum attainable Lorentz factor at present time is less than $\gamma$. 
\begin{equation}
\frac{\gamma_m}{1 + A\gamma_mt_2} < \gamma
\end{equation} 
\begin{equation}
t_2 > \frac{\gamma - \gamma_m}{\gamma\gamma_m}
\end{equation}

So, for calculating $p(\gamma)$ for $\gamma< \gamma_m$, lower limit of the integral in the equation 18 will change to $(\gamma - \gamma_m)/\gamma\gamma_m$ instead of t. So, for $\gamma < \gamma_m$
\begin{equation}
p(\gamma) = \int_{\frac{\gamma - \gamma_m}{\gamma\gamma_m}}^t \frac{K}{t}\frac{\gamma^{-p}}{{(1-A\gamma t)}^{-p+2}} dt_2 \hspace{10 mm} 
\end{equation}
\begin{equation}
p(\gamma) = \frac{K\gamma^{-(p+1)}}{A(p-1)t}\left[\left(\frac{\gamma}{\gamma_m}\right)^{p-1} - \left(1 - \frac{\gamma}{\gamma_c}\right)^{p-1}\right] \hspace{10 mm} \gamma < \gamma_m
\end{equation}

\subsection{Fast Cooling}
The main logic remains the same for fast cooling also, equation 16 and 17 are always valid. If you are looking in the region where $\gamma > \gamma_c$, then, upper limit in the integral in equation 18 should be as given by equation 24 instead of $t$. Similarly if you are looking in the region where $\gamma < \gamma_m$, the lower limit of the integral in equation 18 should not be zero. But, as given by equation 30. Now, in case of fast cooling, if you are looking at the region $\gamma_c < \gamma < \gamma_m$, only those electrons have a non-zero probability of having this Lorentz factor which were injected such that the minimum attainable Lorentz factor at present time is less than $\gamma$ and the maximum attainable Lorentz factor at present time is greater than $\gamma$ i.e. injections which satisfies both equation 26 and 30. So, both the limits of the integral in equation 18 will change according equations 24 and 30.

So, for fast cooling and $\gamma_m < \gamma < \gamma_c$,
\begin{equation}
p(\gamma) = \int_{\frac{\gamma - \gamma_m}{\gamma\gamma_m}}^{\frac{1}{A\gamma}} \frac{K}{t}\frac{\gamma^{-p}}{{(1-A\gamma t)}^{-p+2}} dt_2 
\end{equation}
\begin{equation}
p(\gamma) = \frac{K}{A(p-1)t} \frac{\gamma^{-2}}{\gamma_m^{p-1}} \hspace{10 mm} \gamma_m < \gamma < \gamma_c
\end{equation}
 
For $\gamma > \gamma_m$, only upper limit changes because $\gamma > \gamma_c$ but not $< \gamma_m$, so the expression for $p(\gamma)$ is exactly same as equation 26.

Similarly, for $\gamma< \gamma_c$, only the lower limit changes because $\gamma < \gamma_m$ but not $\gamma > \gamma_c$, so the expression is exactly same as equation 30.
A summary of formulas for $p(\gamma$) in different regimes for both slow and fast cooling case is given below:

\vspace{5 mm}

\emph{Slow Cooling} 

\begin{equation}
p(\gamma) = \frac{K\gamma^{-(p+1)}}{A(p-1)t}\left[\left(\frac{\gamma}{\gamma_m}\right)^{p-1} - \left(1 - \frac{\gamma}{\gamma_c}\right)^{p-1}\right] \hspace{10 mm} \gamma < \gamma_m \\
\end{equation}

\begin{equation}
p(\gamma) = \frac{K}{A}\frac{\gamma^{-p+1}}{(p-1)t}\left[1 - {(1 - \frac{\gamma}{\gamma_c})}^{p-1}\right] \hspace{10 mm} \gamma_m \leq \gamma \leq \gamma_c \\
\end{equation}

\begin{equation}
p(\gamma) = \frac{K}{A(p-1)t}\gamma^{-(p+1)} \hspace{10 mm} \gamma > \gamma_c
\end{equation}

\vspace{5 mm}
\emph{Fast Cooling} 

\begin{equation}
p(\gamma) = \frac{K\gamma^{-(p+1)}}{A(p-1)t}\left[\left(\frac{\gamma}{\gamma_m}\right)^{p-1} - \left(1 - \frac{\gamma}{\gamma_c}\right)^{p-1}\right] \hspace{10 mm} \gamma < \gamma_m \\
\end{equation}

\begin{equation}
p(\gamma) = \frac{K}{A(p-1)t} \frac{\gamma^{-2}}{\gamma_m^{p-1}} \hspace{10 mm} \gamma_m < \gamma < \gamma_c \\
\end{equation}

\begin{equation}
p(\gamma) = \frac{K}{A(p-1)t}\gamma^{-(p+1)} \hspace{10 mm} \gamma > \gamma_c
\end{equation}

\vspace{5 mm}

These are probability distributions, for obtaining particle distribution functions, multiply all of these equations by $N = N_0t$ (total number of particles). It will simply change $K/t$ in all the above equations to $K_e$.
\begin{equation}
N_0t \times \frac{K}{t} = N_0K = K_e \text{(mentioned earlier)} 
\end{equation}

The particle distribution functions are given below:

\emph{Slow Cooling} 

\begin{equation}
n(\gamma) = \frac{K_e\gamma^{-(p+1)}}{A(p-1)}\left[\left(\frac{\gamma}{\gamma_m}\right)^{p-1} - \left(1 - \frac{\gamma}{\gamma_c}\right)^{p-1}\right] \hspace{10 mm} \gamma < \gamma_m \\
\end{equation}

\begin{equation}
n(\gamma) = \frac{K_e}{A}\frac{\gamma^{-p+1}}{(p-1)}\left[1 - {(1 - \frac{\gamma}{\gamma_c})}^{p-1}\right] \hspace{10 mm} \gamma_m \leq \gamma \leq \gamma_c \\
\end{equation}

\begin{equation}
n(\gamma) = \frac{K_e}{A(p-1)}\gamma^{-(p+1)} \hspace{10 mm} \gamma > \gamma_c
\end{equation}

\vspace{5 mm}
\emph{Fast Cooling} 

\begin{equation}
n(\gamma) = \frac{K_e\gamma^{-(p+1)}}{A(p-1)}\left[\left(\frac{\gamma}{\gamma_m}\right)^{p-1} - \left(1 - \frac{\gamma}{\gamma_c}\right)^{p-1}\right] \hspace{10 mm} \gamma < \gamma_m \\
\end{equation}

\begin{equation}
n(\gamma) = \frac{K_e}{A(p-1)} \frac{\gamma^{-2}}{\gamma_m^{p-1}} \hspace{10 mm} \gamma_m < \gamma < \gamma_c \\
\end{equation}

\begin{equation}
n(\gamma) = \frac{K_e}{A(p-1)}\gamma^{-(p+1)} \hspace{10 mm} \gamma > \gamma_c
\end{equation}

\vspace{5 mm}
Figure 2 and Figure 3 shows the variation of particle distribution as a function of $\gamma$ for the slow cooling case and fast cooling case respectively in logarithmic scale.

\begin{figure}[h!]

\includegraphics{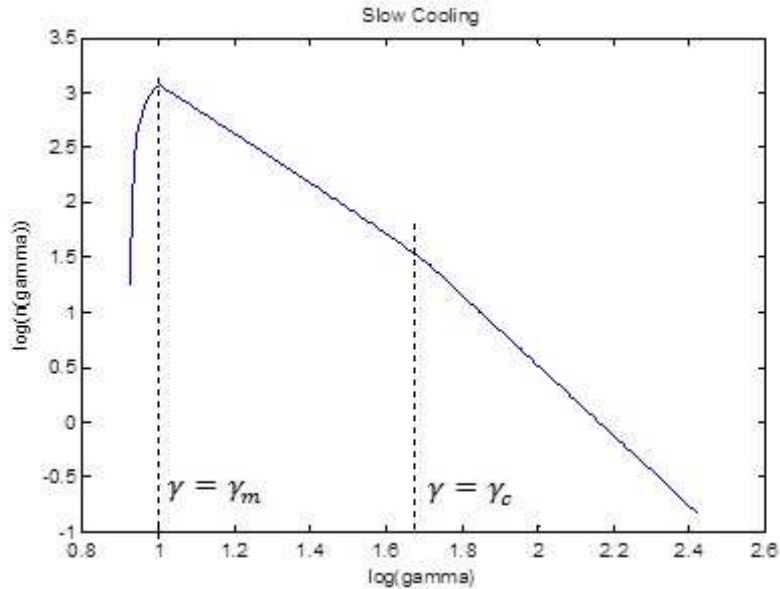}
\caption{Particle Distribution Function for Slow Cooling case  given by equations 42-44}

\end{figure}

\begin{figure}[h!]

\includegraphics{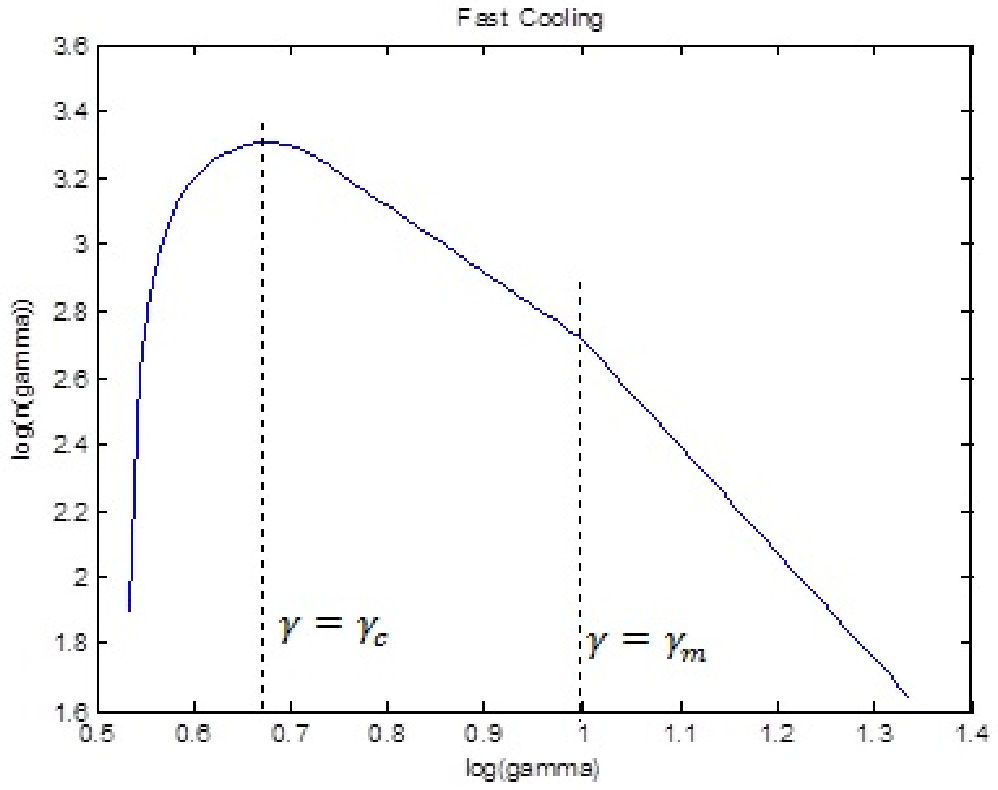}
\caption{Particle Distribution Function for Fast Cooling Case given by equations 45-47}

\end{figure}

\section{Standard Result}
The standard result \citep*{6} obtained by solving the continuity equation says that electron distributions are multi-powerlaws with power indexes defined as follows: 
\begin{itemize}
\item p for $\gamma_m < \gamma \leq \gamma_c$ and p+1 for $\gamma_c < \gamma$
\item -2 for $\gamma_m < \gamma \leq \gamma_c$ and p+1 for $\gamma_c < \gamma$
\end{itemize}
Analytical Solutions exists only in approximate regimes and full result requires numerical calculations.  

\section{Conclusion}

We can see that the results obtained by this theoretical probabilistic model are similar to the ones obtained by solving the continuity equation. $p+1$ power-law index part is exactly same for both slow cooling and fast cooling. Also, $2$ power-law index part is same for fast cooling and for much smaller values of $\gamma$, even equation 43 approximates to a power-law index $p$. However, probabilistic model opens a new dimension, as it is intermediate to the analytical solutions like \emph{Tucker 1978}. The probabilistic approach to this problem has not been attempted so far to the best of author's knowledge. Such an approach is possible in similar problems which involves the evolution of a distribution.

One striking feature observed in these expressions is the steep rising part ($\gamma < \gamma_m$ in Fig. 2 and 3) which is normally not discussed much. The reason for this is that $\gamma_m$ and $\gamma_m^{'}$ are very close and the fraction of electrons is negligible. But, the model discussed here clearly suggests that there will be some electrons between $\gamma_m$ and $\gamma_m^{'}$ which shows a rising particle distribution as we go towards higher $\gamma$. However, this behavior is confined to a very narrow range of $\gamma$. If the magnetic field is very high then this sharp rise becomes significant as the difference between $\gamma_m$ and $\gamma_m^{'}$ will be more and a larger fraction of electrons will lie in this region. Theoretically this behavior should be present in electron distributions evolving under a magnetic field.

These results were obtained assuming a constant magnetic field for simplicity. But, the general concept is very simple and logical. Even for a time varying magnetic fields we can apply exactly the same procedure. The final expressions will change depending on the functional form of $B$ but, all the integrals/integral equations will be same. For e.g. equations 13, 14 and 17 are not going to change under any circumstances. Also, the logic behind changing the limits in different regimes of $\gamma$ for both slow cooling and fast cooling (equation 24 and 28) is also the same however, the final conditions change depending upon what are the minimum and maximum values of $\gamma$ at a later time $t$. 

The main logic behind the approach is that any particle distribution problem can be modeled as a probability problem because both are similar mathematically (just a difference of normalization). In a generic sense, if we have a particle distribution and it is varying with respect to any variable (spatial or temporal), we can model it as a probability problem by normalizing the particle distribution.

\section{Acknowledgments}

I gratefully acknowledge Resmi L., my project guide who constantly appreciated my work and encouraged me to write this article.

\clearpage
\appendix

\section{Appendix material}
Synchrotron Power emitted by an ultra-relativistic electron under a constant magnetic  
field
is given by\citep*{5}
\begin{equation}
P = \frac{4}{3}\sigma_{t}c\beta^{2}\gamma^{2}U_B = \frac{4}{3}\sigma_{t}c\beta^{2}\gamma^{2}\frac{B^2}{8\pi}
\end{equation}
where $U_B$ is magnetic energy density = $frac{B^2}{8\pi}$. Also, we know that:
\begin{equation}
P = -\frac{dE}{dt} = -\frac{d}{dt} (\gamma m_{e}c^{2})
\end{equation}
Substituting Equation 49 in Equation 48:
\begin{equation}
-m_{e}c^{2}\frac{d\gamma}{dt} = \frac{4}{3}\sigma_{t}c\beta^{2}\gamma^{2}\frac{B^2}{8\pi} = \frac{\sigma_{t}c}{6\pi} B^{2}\gamma^2
\end{equation}
Separating the variable $\gamma$ and $t$, and then integrating we get:
\begin{equation}
\int_{\gamma_0}^{\gamma} \frac{d\gamma}{\gamma^2} = \int_{0}^{t} -\frac{\sigma_{t}B^2}{6\pi m_{e}c} dt
\end{equation}
\begin{equation}
\frac{1}{\gamma_0} - \frac{1}{\gamma} = \frac{-\sigma_{t}B^2}{6\pi m_{e}c} t = -At
\end{equation}
\begin{equation}
\gamma-\gamma_0 = -\gamma\gamma_0 At
\end{equation}
\begin{equation}
\gamma(t) = \frac{\gamma_0}{1 + A\gamma_0t} \hspace{10 mm}where\hspace{10 mm} A = \frac{\sigma_{t}B^2}{6\pi m_{e}c} 
\end{equation}



\label{lastpage}

\end{document}